\documentclass[showpacs,twocolumn,preprintnumbers,amsmath,amssymb,superscriptaddress]{revtex4}

\usepackage{amsmath}
\usepackage{amssymb}
\usepackage{amsfonts}
\usepackage{fancyhdr}
\usepackage{fancybox}
\usepackage{color}
\usepackage{graphicx}
\usepackage{float}
\usepackage{amsthm}

\renewcommand{\Im}{{\rm Im}}

\makeatletter

\newcommand{\R}{{\mathbb R}}

\newcommand{\crl}[1]{[-\infty,\infty]}

\newcommand{\ave}[1]{\langle{#1}\rangle}

\newcommand{\av}[1]{\langle#1\rangle}

\usepackage{amsthm}

\begin{document}
\title{Reply to ``Comment on `Optimal probe wave function of weak-value amplification' "}
 \author{Yuki Susa}
 \email{susa@th.phys.titech.ac.jp}
 \affiliation{Department of Physics, Tokyo Institute of Technology, Tokyo, Japan}
 \author{Yutaka Shikano}
 \email{yshikano@ims.ac.jp}
 \affiliation{Institute for Molecular Science, Okazaki, Aichi, Japan}
 \author{Akio Hosoya}
 \email{ahosoya@th.phys.titech.ac.jp}
 \affiliation{Department of Physics, Tokyo Institute of Technology, Tokyo, Japan}
\date{\today}
\begin{abstract}
It is pointed out that the ``counter example" presented in the Comment is a family of probe wave functions which are increasingly
broad as the shift becomes large. Furthermore, the author's variational calculation is not correct in the sense that we have 
to gauge fix the freedom of the phase translation. It is shown that there are two 
kinds of solutions, normalizable and un-normalizable. The former is our optimal solution, and the latter is what he found. 
It seems only the former is relevant from a practical point of view.
 \end{abstract}
\maketitle
The Comment written by Di Lorenzo~\cite{ADL} is nice for critically looking at the same problem from different angles. 
This would not only clarify the points but also would shift the emphasis to more appropriate ones than presented in our original 
paper~\cite{SSH}. His Comment is in this category, although we disagree with his conclusions. Throughout our Reply, 
the notation is used in Ref.~\cite{ADL}.

We agree that the trial function (3) in Ref.~\cite{ADL} gives an arbitrarily large shift in the position in proportion 
to the parameter $\alpha$. However, the variance in the position becomes much larger in proportion to 
the parameter $\alpha^2$. Namely, the probe wave function becomes broader as its center becomes large. 
In our final optimal probe wave function (18) in Ref.~\cite{SSH}, the variance becomes 
infinity except for $\av{\hat{x}}_{f}=2 m$ ($m \in \mathbb{Z}$). Therefore, both cases 
are practically useless because of the large variance while we obtain the large shift. 
It is also noted that the shift on his trial function can exceed our optimal shift 
since our optimal solution gives the stationary value for the shift optimality.

However, in this exceptional case, $\av{\hat{x}}_{f}=2 m$ ($m \in \mathbb{Z}$), 
our final optimal probe wave function with normalization is the Kronecker $\delta$ as
\begin{align}
\tilde {\xi}_{f}(x = 2 n ) &= \frac{2}{\pi}\frac{\sin \left[ \frac{\pi}{2}(2n-2m)\right]}{2n-2m} \notag \\ 
&= \frac{\sin \left[ (n-m) \pi \right]}{(n-m)\pi} \notag \\ 
& =\delta_{mn} \ \ \ (n \in \mathbb{Z}).
\end{align}
Therefore, this gives zero variance. As mentioned in the main text of Ref.~\cite{SSH}, it is remarked 
that the domain of our probe wave function in the momentum representation is finitely bounded
\begin{equation}
	\left| k \right| \leq \pi / 2,
\end{equation}
with  
\begin{equation}
	\left| \xi_i (k = \pi / 2) \right|^2 = \left| \xi_i (k = - \pi / 2) \right|^2\hspace{-0.2em},
\end{equation}
which implies the periodic boundary condition for the wave function $\xi_{i} (k)$ up to phase
so that the position operator $\hat{x}$ has a discrete spectrum for our final optimal probe wave function. 

To obtain the sharp final distribution with the large shift, the particular case, 
$\av{\hat{x}}_{f}=2 m$ ($m \in \mathbb{Z}$), is important, which can be achieved 
by tuning the pre- and postselected states, i.e., the weak value.
We believe that the probe wave function is more practical if the shift is larger while the variance is smaller, 
although we do not rule out the usefulness of a broad wave function of the large shift.

As he correctly remarked, the quantity $\Delta \ave{\hat{x}}$ is invariant under the operation $\xi_i \rightarrow e^{ix_{0}k} \xi_i$ 
for any $x_0 \in \R$. Due to this gauge symmetry, the functional $\Delta \ave{\hat{x}}$ is flat in this phase shift direction 
in the function space so that 
it does not have a local extremum. A standard method to cope with this problem is the gauge fixing. 
More precisely, the difference $\av{\hat{x}}_{f}-\av{\hat{x}}_{i}$ is gauge invariant and coincides with $\av{\hat{x}}_{f}$ by imposing gauge fixing condition $\av{\hat{x}}_{i}$.

To do this, we add a Lagrange multiplier term,
\begin{equation}
 - \Im \left[ \mu \int dk\,\xi^{*}_{i}\xi^\prime_{i} \right],
\end{equation}
where $\mu$ is the Lagrange multiplier. 
Upon the variation with respect to $\mu$, we obtain $\ave{\hat{x}}_i=0$.

Then, Eq. (A4) in Ref.~\cite{ADL} should be replaced by 
\begin{equation}
\Delta \ave{\hat{x}}= - \Im \left\{ \frac{\int dk [B\xi_{i}]^{*}[B\xi_{i}]^{\prime}}{N_f} - \tilde{\mu} \int dk\,\xi^{*}_{i}\xi'_{i} \right\},
\end{equation}
where the combination $\tilde{\mu} := \mu+1/N_i$ can be taken as an independent variable.

The rest of the calculation is straightforward, and Eq. (A6) in Ref.~\cite{ADL} is reproduced except that  the quantity $|\bar{B}|^2$ is replaced by 
$\tilde{\mu}N_f$ and $\ave{\hat{x}}_i =0$. If  $\tilde{\mu}\neq 0$, the solution becomes un-normalizable as the author pointed out. 
However, if $\tilde{\mu}=0$, our optimal solution (15) in Ref.~\cite{SSH} is reproduced, which is normalizable.

We think that only the normalizable and stationary solution for the probe wave function is relevant 
so that the case $\tilde{\mu}\neq 0$ is excluded. The point is the gauge fixing, which makes the multiplier variable. 

The authors thank A. Di Lorenzo for valuable discussions. This work is partially supported by the Global Center of Excellence
Program ``Nanoscience and Quantum Physics" at the Tokyo Institute of Technology.

\end{document}